\documentclass[
	a4paper, 
	10pt, 
	unnumberedsections, 
	twoside, 
]{LTJournalArticle}

\runninghead{} 

\footertext{\textit{RISC-V Summit Europe, Paris, 12-15th May 2025}}  

\title{RISC-V for HPC: An update of where we are and main action points} 

\author{
	Nick Brown\textsuperscript{1}\thanks{Corresponding author: \href{mailto:n.brown@epcc.ed.ac.uk}{\tt Nick Brown (n.brown@epcc.ed.ac.uk)}}
}

\date{\footnotesize\textsuperscript{\textbf{1}}EPCC, The University of Edinburgh, 47 Potterrow, Edinburgh, UK}

\renewcommand{\maketitlehookd}{%
	\begin{abstract}
This extended abstract is submitted on behalf of the RISC-V HPC SIG who have been undertaking an analysis to explore the current state and limitations of the RISC-V ecosystem for HPC. Whilst it is right to celebrate that there has been great progress made in recent years, we also highlight limitations and where effort should be focussed.
	\end{abstract}
}

\begin{document}

\maketitle 


\section{Introduction}

High Performance Computing (HPC) involves the use of computers to simulate the real-world. Given that every major scientific discovery in the past 30 years has, to some extent, involved HPC supercomputers are a critically important resource. As we move further into the exascale era (the ability to perform a billion billion calculations per second), the HPC community is undergoing a transition as it places a greater emphasis on sustainability of operations.

Whilst RISC-V has grown rapidly in areas such as embedded computing, it is yet to gain significant traction in High Performance Computing (HPC). It is not enough for RISC-V to be \emph{just as good} as existing technologies commonly used in HPC, such as x86 CPUs coupled with Nvidia or AMD GPUs, but instead it must provide capabilities over and above incumbents in order to displace them.

\subsection{Why RISC-V in HPC?}

The main metrics that the HPC community cares about are performance, availability of hardware, energy efficiency, cost, and risk. Put simply, to drive adoption of RISC-V in HPC vendors should aim to maximimise the first three whilst minimising the last two, and it is the role of the RISC-V community and HPC SIG to assist here and make a strong case. 

There is a strong argument that opportunities for specialisation offer significant potential for driving performance, energy efficiency and cost, for example \cite{brown2024accelerating} demonstrated that the Tenstorrent Grayskull delivered comparable performance to a 24-core Xeon Platinum CPU for a scientific computing workload whilst requiring five times less energy and at a considerably lower price point. 

Whilst RISC-V workshops and events organised by the HPC SIG co-located at global HPC conferences have proven popular and are growing, reticence from the HPC community often centres around hardware availability and the current maturity of the ecosystem for HPC. To address the former we are seeing increased availability of RISC-V hardware that is a more serious contender for high performance workloads, and the purpose of our analysis is to help address the later by identifying high priority areas for improvement and solve these.

\section{The state of RISC-V for HPC}

\subsection{Application support}

One of the key questions posed by the HPC community is how easily common HPC simulation codes port to RISC-V and what the likely performance will be. Based upon data gathered from the world's supercomputers we have explored the most popular HPC applications, libraries and benchmarks on RISC-V. We will highlight this further in the talk, but in the main, we found that the majority of these built without issue on RISC-V CPUs and could run common use-cases.

Whilst support for vectorisation by RISC-V hardware and compilers is often cited as being of major concern, and indeed this is an obvious area to consider, with current generations of RISC-V hardware we have found that there are other performance limitations which are potentially more severe. For example, the 64-core SG2042 is a potential contender for HPC and in \cite{brown2025performance} it was found that for compute bound workloads (the NPB EP benchmark) the CPU performs rather well, for example matching the performance core-for-core of a Marvel ThunderX2 which has the same width vectorisation and the large core count of the SG2042 means that when running at larger numbers of cores this CPU outperforms x86 server grade CPUs that have fewer cores. However, for workloads that are either memory bandwidth or latency bound, the SG2042 struggles significantly against other server class CPUs, even at the higher core counts falling behind these other CPUs at much lower numbers of cores. Whilst Sophon might not have placed as much importance on the performance of the memory subsystem for the SG2042 based upon their main markets, this is crucial for HPC and demonstrated in \cite{brown2025performance} that it is a limit.

Of course, porting codes to RISC-V accelerators such as hardware from Esperanto or Tenstorrent inevitably requires code modification as is the case with porting to GPUs. These vendors should consider supporting common pragma based approaches, e.g. OpenMP and OpenACC via C, C++ and Fortran.

\subsection{Software tooling}

GCC and LLVM compiler support for RISC-V is fairly mature. There is a common belif in the community that Fortran compiler support for RISC-V lags other architectures, and as such we undertook an experiment that used NASA's parallel benchmark suite, NPB, to compare timing between each benchmark written in Fortran and C. Table \ref{tab:npb_perf} illustrates the number of times faster the Fortran version is than the C version on both RISC-V and x86. The purpose of this is to compare the numbers and see whether Fortran vs C is much faster on x86 than RISC-V, which would indicate that there are issues with RISC-V Fortran support. Indeed, we do not see this in Table \ref{tab:npb_perf} and the numbers are fairly comparable (apart from the BT benchmark, where the Fortran compiler is very much faster than the C on RISC-V). However, one general limitation is that whilst the MLIR compiler framework provides x86 vectorisation, Neon and SVE dialects, there is not a RISC-V vectorisation dialect. Previous work explored such a dialect in 2022, and a recommendation for the community is to pick this up and mature it.

\begin{table}[htb]
    \centering
    \caption{Speed up of Fortran vs C for NPB benchmarks (class C) for GCC v13 on single core of an SG2042.}
    \label{tab:npb_perf}
    \begin{tabular}{|c|c|c|}
    \hline
      \textbf{Benchmark} & RISC-V & x86 (AMD Rome) \\
     \hline
	  BT & 3.35 & 1.30\\        
	  CG & 1.01 & 0.95\\
        EP & 2.45 & 2.95\\
        FT & 1.47 & 1.13\\
        LU & 2.15 & 1.38\\
    \hline
    \end{tabular}
\end{table}

Where RISC-V does fall short on is performance analysis tooling, this is critical for HPC application developers and currently the only option is to use the Extrae tool from BSC. Whilst this is a decent tool, HPC application developers expect a range of performance tools and-so an important action point is to encourage others to port their tooling, such as Tau, to RISC-V. Indeed it is also important to engage with the vendors and influence them to provide adequate performance counters, for example whilst performance counters are provided by the SG2042 there are several commonly required for performance analysis missing.

\subsection{Infrastructure}

There are several RISC-V testbeds that have been stood up with the intention of providing an HPC like service. Indeed, infrastructure to support this such as the module environment and Slurm batch submission system is already present and working well on RISC-V, providing users with a familiar experience. These systems currently use NFS for a shared filesystem, which again works well on RISC-V but there is work to be done to mature and optimise high performance parallel filesystems, such as Lustre and GPFS in order to be a realistic proposition for large-scale HPC.

However, there are omissions on the systems side where tooling frequently used by system administrators to monitor and manage their systems in the data centre still requires support for RISC-V. Much of this is provided by system integrators, such as HPE, and-so closer coordination with them would be beneficial.

A major limitation of RISC-V for HPC at the moment is the lack of support for high performance networking, making RISC-V an unrealistic proposition for distributed memory parallelism. Whilst there have been some early explorations here, for instance a prototype Infiniband driver for RISC-V, this is very early on and not currently deployable. Mature support for high performance networking technologies should be seen as high priority by the community, because until we address this RISC-V will always be limited when it comes to large-scale HPC deployment.

\section{Conclusions}
The advances in the RISC-V ecosystem have been phenomenal in the past couple of years ago, but there is still a way to go in order to make this a realistic proposition for large-scale HPC. The purpose of this presentation will be to demonstrate where the RISC-V community can contribute and that it is important that RISC-V and the HPC community continue to work together to not only make a strong case for the role of RISC-V in HPC but also to address the blockers that we have identified.





\printbibliography 

@inproceedings{brown2024accelerating,
  title={Accelerating stencils on the Tenstorrent Grayskull RISC-V accelerator},
  author={Brown, Nick},
  booktitle={SC24-W: Workshops of the International Conference for High Performance Computing, Networking, Storage and Analysis},
}

@inproceedings{brown2025performance,
  title={Performance characterisation of the 64-core SG2042 RISC-V CPU for HPC},
  author={Brown, Nick},
  booktitle={International Conference on High Performance Computing},
}


\end{document}